\documentclass[sigconf]{acmart}
\usepackage{epsfig,endnotes}
\usepackage{listings}
\usepackage{tcolorbox}
\usepackage{amsmath}
\usepackage{caption,subcaption}

\definecolor{dkgreen}{rgb}{0,0.6,0}
\definecolor{gray}{rgb}{0.5,0.5,0.5}
\definecolor{mauve}{rgb}{0.58,0,0.82}
\usepackage{enumitem}
\setlist[itemize,1]{leftmargin=10pt}
\setlist[enumerate,1]{leftmargin=12pt}
\setlist[itemize]{topsep=0pt}
\setlist[enumerate]{topsep=0pt}

\usepackage[font=small,labelfont=bf,justification=justified]{caption}

\lstdefinelanguage{nvm}
{morekeywords={void, object, int, long, byte, string, new},
    sensitive=false,
    morecomment=[s]{@}{_},
    morecomment=[s]{/*}{*/},
}

\setcopyright{none}
\renewcommand\footnotetextcopyrightpermission[1]{} 
\makeatletter
\renewcommand\@formatdoi[1]{\ignorespaces}
\makeatother
\acmDOI{}

\acmISBN{}

\acmConference[]{}{}{} 
\acmYear{2018}
\copyrightyear{2018}

\begin{document}
%\sloppy

%don't want date printed
\date{}

%\doi{nnnnnnn.nnnnnnn}
% \title{\Large \bf Application Aware Tiered Object Storage Layout \\
% using Non-Volatile Memory}
\title{Tiered Object Storage using Persistent Memory}
% * <amisaha@cisco.com> 2018-01-23T06:16:53.477Z:
%
% ^.

%for single author (just remove % characters)
\author{
{\rm Johnu George, Ramdoot Pydipaty, Xinyuan Huang, Amit Saha, Debojyoti Dutta}\\
%{\rm J. George, R. Pydipaty, X. Huang, A. Saha, D. Dutta}\\
Cisco Systems
\and
{\rm Gary Wang, Uma Gangumalla}\\
Intel Corporation
% copy the following lines to add more authors
% \and
% {\rm Name}\\
%Name Institution
} % end author

%\authorinfo{Mr. and Mrs. Double Blinded}
%           {Paper Id. 196}
%           {Pages: 11}

\maketitle

% Use the following at camera-ready time to suppress page numbers.
% Comment it out when you first submit the paper for review.
%\thispagestyle{empty}

\subsection*{Abstract}
Most data intensive applications often access only a few fields of the
objects they are operating on. Since NVM provides fast,
byte-addressable access to durable memory, it
is possible to access various fields of an object stored in NVM directly without
incurring any serialization and deserialization cost.
%However, NVM is
%expensive, slower than DRAM, and cannot scale to the size of HDDs/SSDs,
%thus warranting a usage
%model that can increase the utilization of NVM.
This paper proposes
{\it a novel tiered object storage model} that modifies a data structure
such that only a chosen subset of  fields of the data structure are
stored in NVM, while the remaining fields are stored in a cheaper
(and a traditional) storage layer such as HDDs/SSDs.
We introduce a novel {\it linear-programming based optimization framework}
for deciding the field placement.
Our proof of concept demonstrates
that a tiered object storage model improves the execution
time of standard operations by up to 50\% by avoiding the cost of
serialization/deserialization and by reducing the memory footprint of
operations. 
%We believe that this paper will lead to several new directions in accelerating data intensive 
%workloads in a tiered memory-storage setup. 

\section{Introduction}
\label{introduction}

Persistent Memory (PMEM)~\cite{acmq:nvm}, also known as Non Volatile Memory
(NVM) or Storage Class Memory (SCM) is one of the disruptive trends
in the compute technology landscape. Unless mentioned otherwise, this
article uses NVM, NVRAM, NVDIMM, pmem, persistent memory, and storage class memory
interchangeably.
In addition to {\it providing data durability}, these memories are
{\it byte addressable} and have access speeds comparable to that of DRAM.
%than traditional
%secondary storage technologies (including traditional NAND Flash).
For example, the imminent dual in-line non-volatile memory (NVDIMM) from 
Intel, 3D-XPoint~\cite{3dxpoint_dimm}, touts an access latency
of around 500~nanosecond (ns),
i.e., within an order of magnitude of DRAM (100~ns) and much faster than
the 30,000~ns for NVMe-SSDs, which are simply faster {\it block devices}.

\if 0
We believe that access to byte-addressable, persistent memory is going to be
a dominant trend in
the compute industry that will allow applications to directly access memory
instead of serializing/de-serializing to/from a block device. 
The folklore of I/O being
considerably slower than computation is being questioned with the introduction
of NVDIMM devices. This
realization has already led the research community to focus on {\it memory
driven computing}~\cite{fast17:keeton}
But, persistent memory is not a replacement for DRAM
or storage devices like HDDs and SSDs because it is not as fast as DRAM and
cannot yet scale to the size of HDDs/SSDs. 
\fi

Most applications use few fields of entire objects during
computation. For example, several real-time data logging applications
are interested only in a few (3--5) fields embedded in log lines.
Similarly, large graph processing applications
often access a small subset of the graph structure. For
example, Facebook's graph API~\cite{facebook_graph:url} returns more than {\it
fifty} fields in response to a query for a user. But, an algorithm that is
looking for, say, {\it connections} in similar age group or living in the same
geographical area of a specified user, {\it need not} access
the all fields of the object representing the user.
\if 0
To motivate further, let us consider a test case wherein a
file (on disk) contains a collection of $N$ objects,
where each object has several fields, such as ``age'',
``name'', and ``image''.
Suppose we want to find the age of a person named ``Bob''.
To solve the problem, we only require two fields ---
``age'' and ``name'' fields from all objects.
Since disk is a block device, the entire file has to be read
from disk and {\it deserialized} to read the relevant data. An iterative
search routine accesses all $N$ objects to return the ``age'' of the
object with ``name'' as ``Bob''. The situation worsens if all the $N$
objects do not fit in memory, leading to swapping, 
thus further increasing the processing time.
\fi
\if 0
The basic issue is that {\it the current scenario demands
the entire file to be processed even if it is only required to access few fields
out of the entire object structure}. 
\fi

In our proposal, objects are no longer considered a single entity in any
layer (unlike caching system). Fields of the objects that can stay volatile are
kept in DRAM, whereas fields that need persistence, are either kept in pmem,
or on disk in serialized form.
%The logic for choosing appropriate storage for the various
%fields of an object will be discussed in the later section.
%In the previous
%example, the ``name" and ``age" fields from all objects can be stored in
%pmem,
%while the remaining fields can be on disk.
This reduces main memory usage and
fits more objects in pmem. This leads to more efficient use of pmem space
since pmem is more expensive than DRAM.
Additionally, since other fields are not
needed for the computation, this avoids disk access, thus speeding up the
processing.
In case of garbage collected languages such as Java, 
this reduction of memory usage leads to less frequent garbage collection triggers,
directly improving the running time of applications.

\subsection*{Contributions}
%\noindent
%{\bf Contributions:}
The contributions of this paper are as follows:
\begin{enumerate}
\itemsep0em 
\item A novel tiered storage layout that allows distributing individual 
fields of a single data structure into multiple storage layers at run time.
%\item We show that manual tagging of selectively placing fields in persistent memory
%leads to 50 percent more efficiency when we use tiered storage layouts. 
\item A novel technique called {\it profiled tagging}
where the results of benchmarking applications are fed into
a linear-programming optimization framework to
determine fields to be stored in pmem given the system and cost characteristics.
\item Finally, our evaluation on $k$-means clustering and
graph search demonstrates that the proposed model indeed
improves performance considerably.
\end{enumerate}

The rest of the paper is organized as follows.
In Section~\ref{related} we differentiate our work from prior related work.
We present the proposed tiered storage design in Section~\ref{design}.
We present our evaluation in
Section~\ref{evaluation} and finally, we conclude in Section~\ref{future}.

\section{Related Work}
\label{related}
The existing work in providing a file system access to
persistent memory, such as
BPFS~\cite{condit:sosp09}, PMFS~\cite{dulloor:eurosys14}, and
SCMFS~\cite{wu:sc11} are irrelevant to our work.
The NVM based file system by Wei et al.~\cite{wei:icpads15},
keeps meta-data in NVM 
(making meta-data access fast and persistent)
while keeping the actual objects in
slower hard disks. This is an example of a tiered file system
but is not extensible to individual data structures,

Unlike this proposal,
data structures specifically designed
for NVM such as, NV-heaps~\cite{coburn:asplos11},
Write Optimized Radix Tree (WORT) \cite{lee:fast17}, and
Consistent and Durable Data Structure (CDDS)~\cite{venkataraman:fast11},
do not consider splitting up individual data structures.
Malicevic et al.~\cite{malicevic:inflow15} intelligently
%place application data in a hybrid NVM and DRAM system. 
%This work puts
certain strategic data structures in DRAM (as opposed to NVM) to
avoid the delay of accessing NVM.
Similarly, Dulloor et al.~\cite{dulloor:eurosys16}
store some data structures in DRAM and the rest in NVM, with the
goal of avoiding application slowdown by moving data structures out of
DRAM to NVM. Unlike this proposal, these works do not address the {\it real
possibility of splitting up individual data structures by storing different
fields in different types of memory}.

\begin{figure*}[h]
%     \begin{subfigure}{0.33\textwidth}
%         \centering
%         \includegraphics[width=\linewidth]{figs/dram.png}
%         \caption{Simplified native Java}
%         \label{fig:native_java}
%     \end{subfigure}
     \begin{subfigure}{0.45\textwidth}
        \centering
        \includegraphics[width=\linewidth]{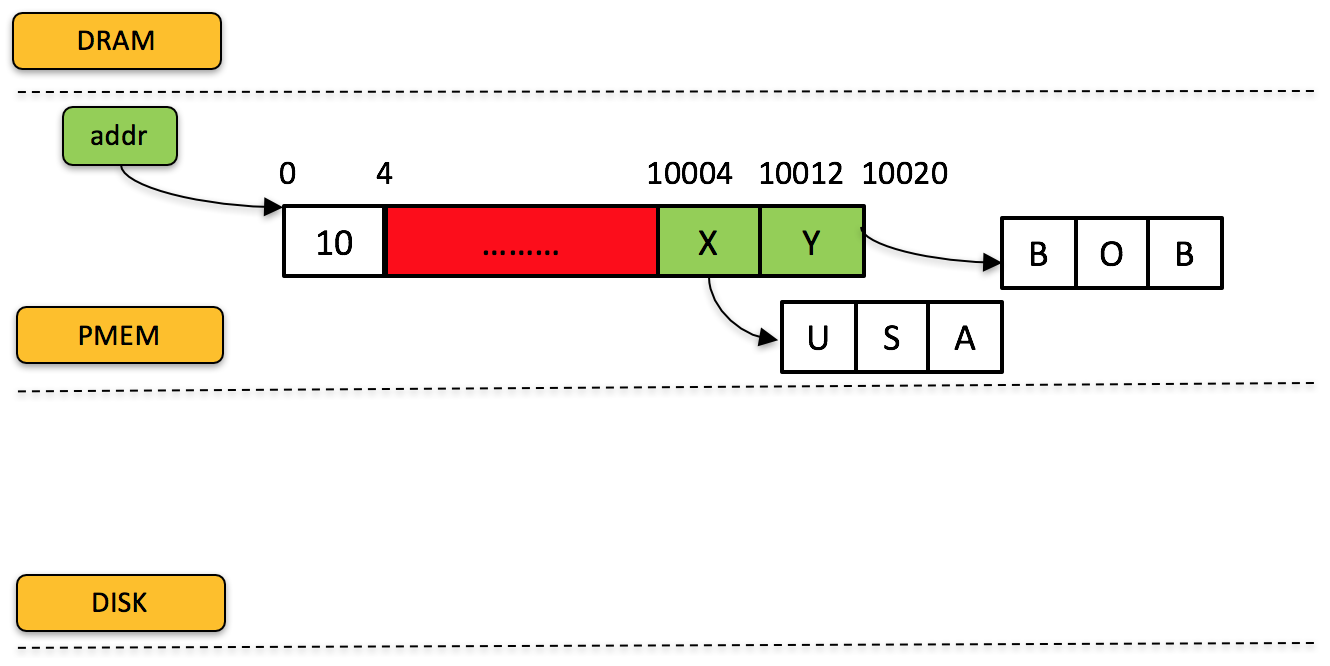}
        \caption{DRAM and pmem}
        \label{fig:dram_pmem}
    \end{subfigure}
    \begin{subfigure}{0.43\textwidth}
        \centering
        \includegraphics[width=\linewidth]{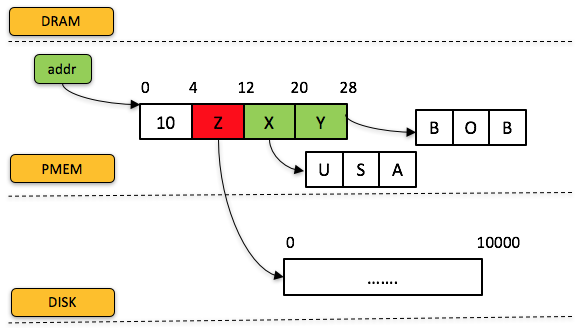}
        \caption{DRAM, pmem, and disk}
        \label{fig:dram_pmem_disk}
    \end{subfigure}
    \caption{Partitioning of data object among various locations.
    The ``age" field starts at byte offset 0 and the ``image" field starts
    at byte offset 4 from the object start address (assuming
    the Integer size is 4 bytes). Variable sized fields are stored via
    indirections whereas fixed sized fields are stored directly.}
\label{fig:object_partition}
\end{figure*}

Mnemosyne~\cite{mnemosyne:asplos11} is an interface for creating, managing, and
maintaining consistency in data structures stored on persistent memory. This
work introduced the concept of user defined annotation ({\it pstatic}) that allows
the system to identify which objects are to be persisted. We
use a similar annotation for manual tagging of data structure
fields (Section~\ref{manual_tag}). But, that is where the similarity ends.

Wei et al.~\cite{wei:icpads15}
proposed a NVM based file system that keeps meta data in NVM (thus
making meta data access fast and persistent) while keeping the actual objects in
slower hard disks. This is an example of a tiered storage system but from a
file system point of view. It is not extensible to individual data structures,
which is the focus of the work presented in this paper.

Alluxio~\cite{alluxio:url} uses cache tiering, assuming that
tiers are ordered from top to bottom based on I/O performance. This top to
bottom hierarchy is not valid anymore with NVM since NVM
has hybrid features of both durability and byte addressability.
%Moreover, systems like Alluxio are block storages
%without any preference for a specific data layout.
To the best of our knowledge, this is the first proposal to split
individual data structures and store the different fields in
different tiers of a hybrid, tiered storage system.

Andrei et al.~\cite{sap_hana:vldb2017} modified the SAP HANA,
a column-oriented RDMS, to include NVRAM.
They specifically put the {\it Main Column Fragments} in NVRAM because they
did not want to fundamentally change the structure of the DBMS because that would
make the changes less likely to be adopted in production. Their changes are 
specific to SAP HANA and they do not suggest splitting up
individual data structures.

The idea of splitting up a data structure and storing them into different types 
of memory has some similarity to the concept of
{\it Binary Large OBjects} (BLOBs) that is commonly used in databases for storing
large binary objects, such as images and binary files. BLOB is an
implementation trick and it could be envisioned that a similar technique can be 
used to implement the proposal made in this paper. Our approach, as described next,
is different.

\section{Design}
\label{design}

Though the overall idea is applicable to other languages, 
we base our design on Java because it is one of the
most popular programming languages~\cite{java_projects:url}
used in many popular big data frameworks like
Apache Hadoop~\cite{hadoop}, Apache Spark~\cite{spark},
and Apache Flink~\cite{flink}.

\subsection{Tiered Storage Layout}
\label{fixed}
Let us consider a traditional object, \texttt{person},
%(Listing~\ref{traditional_ds}) 
in which the ``age'', ``place'',
and ``name'' fields are most frequently accessed (by say a search program)
whereas the ``image'' field is only retrieved if the object matches a search
and the person's image has to be displayed. A traditional object does not have
the annotations ``@pmem'' or ``@disk'' as shown in Listings~\ref{image_pmem}
and \ref{image_disk}.
Traditionally, in spite of the ``image'' being seldom accessed, the entire object
has to be brought into heap (DRAM) when the object is searched.
Since we are presenting in a Java context, unless 
otherwise mentioned, ``DRAM'' and ``heap'' are used interchangeably.

{\it Tiered storage layout} uses {\it fixed sized record format} so that
the field offsets are based on the field type. Primitive data types and arrays
have fixed storage sizes (e.g., short -- 2~bytes, int -- 4~bytes, long -- 8~bytes),
while variable sized containers, including {\it Strings},
are stored indirectly via a {\it long}
value that points to a buffer containing
the actual contents of the variable.

\begin{lstlisting}[language=nvm, caption=Annotated object with all fields in pmem, label=image_pmem
%, belowskip=-0.8\baselineskip
]
object person {
    @pmem_ int age;
    @pmem_ byte[10000] image;
    @pmem_ string place;
    @pmem_ string name;
}
\end{lstlisting}
\begin{lstlisting}[language=nvm, caption=Annotated object with just ``image" on disk, label=image_disk]
object person {
    @pmem_ int age;
    @disk_ byte[10000] image;
    @pmem_ string place;
    @pmem_ string name;
}
\end{lstlisting}

Traditional object definitions
%(Listing~\ref{traditional_ds}) 
are annotated as shown in Listing~\ref{image_pmem}
(all fields, including ``image'', are stored in pmem) and
in Listing~\ref{image_disk} (only ``image'' is stored in disk and others are stored
in pmem). This annotation can be done either by manual tagging (Section~\ref{manual_tag})
or profiling (Section~\ref{profile_tag}). 
%\footnote{
%Annotating fields of a class to suggest data location during run time.
%The annotations are before the data types so that a semi-colon (`;')
%still separates one field from the next.}
The actual data for variable sized field is stored
separately while its address is stored in the parent object field. 
The difference in the object layout is depicted in Figure~\ref{fig:object_partition}.
%Figure~\ref{fig:native_java} shows {\it a simplified version} of the traditional object
%layout in which all fields are kept in DRAM (Java heap in case of Java).
Figure~\ref{fig:dram_pmem}
corresponds to Listing~\ref{image_pmem} and shows an object layout in which all fields
are stored in pmem thus allowing the encompassing object to be persistent (unlike
a traditional DRAM based model). Figure~\ref{fig:dram_pmem_disk} corresponds to
Listing~\ref{image_disk} and shows a layout in which
the ``image'' field is stored in disk whereas
the rest are stored in NVM. This saves valuable NVM space by keeping only
{\it the frequently accessed} fields in NVM.
%while relegating the less frequently accessed ``image'' field to disk.

\if 0
In the examples described in Fig.~\ref{fig:dram_pmem}
and Fig.~\ref{fig:dram_pmem_disk}, ``age" field starts at byte offset 0
and ``image" field starts at byte offset 4 from the object start address (assuming
the Integer size to be 4 bytes). The application user can specify the location
of each field where it is stored. Based on the application requirements, the
user can annotate to store the image field
in persistent memory (Listing~\ref{image_pmem})
or on disk (Listing~\ref{image_disk}) thus saving persistent memory space.
\fi

\begin{figure}
	\centering
    \includegraphics[width=0.9\linewidth]{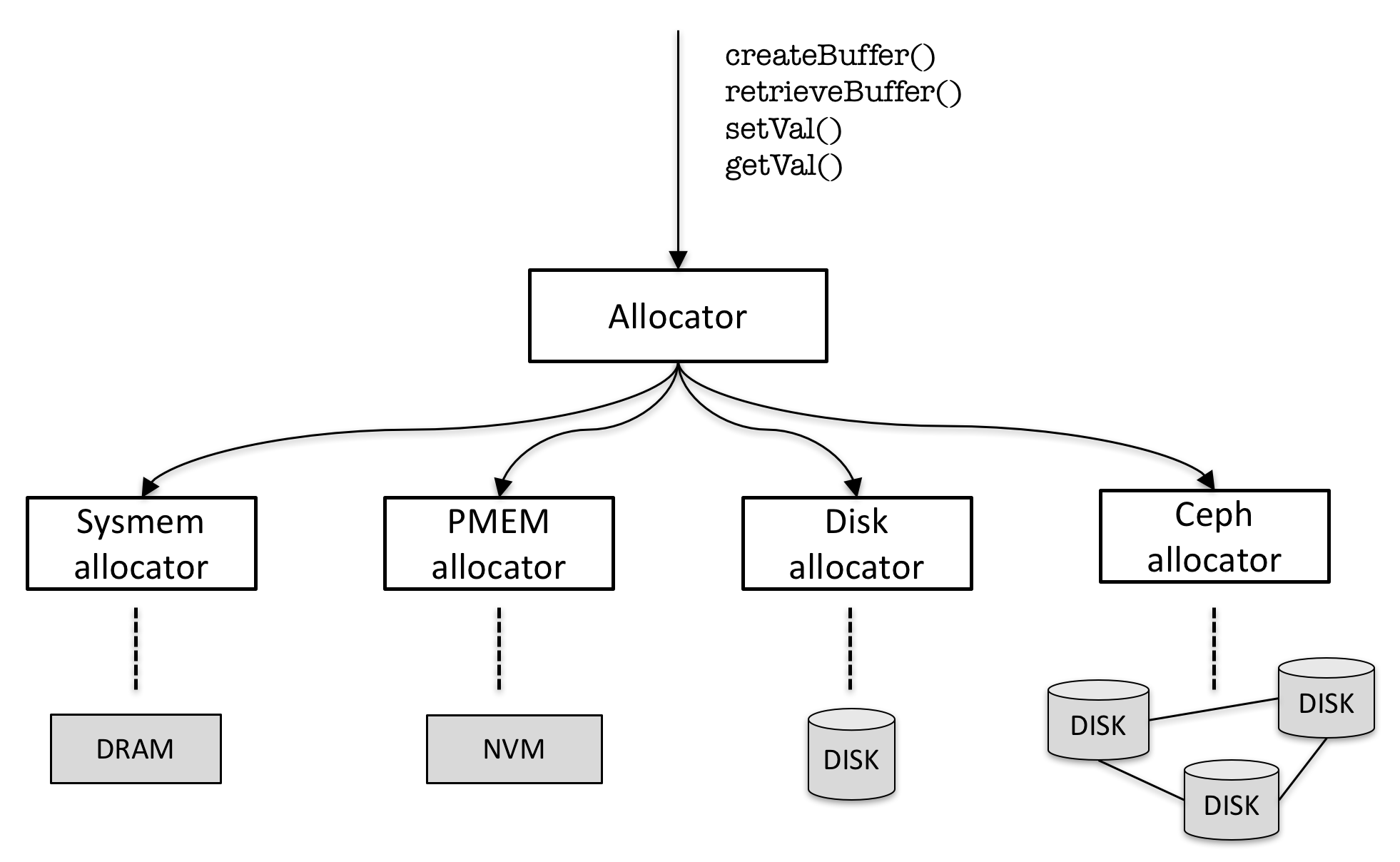}
    \caption{Different types of allocators behind a generic API set.
    The specific names of the APIs
    %such as \texttt{createBuffer}
    %and \texttt{retrieveBuffer}
    would be implementation dependent.
	For persistent memory, libraries such as
    {\it libpmemobj}~\cite{pmem:pmdk} can be used to implement
    memory allocation and transaction semantics on the
    underlying non-volatile memory.
    }
    \label{fig:allocator}
\end{figure}

\if 0
%Finally, to use this {\it durable} version of the \texttt{person} object, the application
%has to use the {\it generated} class, \texttt{newPerson}, as shown in Listing~\ref{allocator}.
%Apart from using a new class name, the application also has to use {\it generated}
%storage GET/SET APIs, as shown in Listing~\ref{generated_apis}.
\fi

\subsection{Generic Storage API}
\label{api}
In order to be compatible with a large set of underlying storage devices,
each device type is modeled as a storage allocator that implements GET/SET
APIs to read and write to
the corresponding devices (Figure~\ref{fig:allocator}). 
There are allocators
for system memory, persistent memory, HDD, SSD, and distributed file
systems like Ceph. Each storage allocator abstracts the underlying device
capabilities hiding the details
of its back-end implementation and implements GET/SET APIs to read and write to
the corresponding device.

\begin{lstlisting}[language=nvm, caption=Generated APIs for ``person'' class of Listing~\ref{image_disk}. PmemAllocator is the allocator for read/write to pmem and DiskAllocator is the allocator for disk.,
					label=generated_apis]
class DurablePerson {
...
/* 'addr' is start address of person object */
void setAge (int age) {
    PmemAllocator.setVal(addr + 0, age);
}
int getAge() {
	return PmemAllocator.getVal(addr + 0);
}
void setImage (byte[] image) {
    Z = DiskAllocator.createBuffer(image);
    PmemAllocator.setVal(addr + 4, Z);
}
byte[] getImage () {
	Z = PmemAllocator.getVal(addr + 4);
    return DiskAllocator.retrieveBuffer(Z);
}
void setPlace (string place) {
    X = PmemAllocator.createBuffer(place);
    PmemAllocate.setVal(addr + 12, X);
}
string getPlace () {
	X = PmemAllocator.getVal(addr + 12);
    return PmemAllocator.retrieveBuffer(X);
}
...
}
\end{lstlisting}

\if 0 % moved to storage.tex
\begin{lstlisting}[language=nvm, caption=Using the generated durable class, label=allocator]
DurablePerson newPerson = new DurablePerson();
/* addr = newPerson.handler();
   addr is internally managed */
newPerson.setAge(10);
newPerson.setImage(byte[1000]{...});
newPerson.setPlace("USA");
newPerson.setName("BOB");
\end{lstlisting}
\fi

As shown in Listing~\ref{generated_apis}, the \texttt{setImage()} function uses a
disk allocator (since Listing~\ref{image_disk} annotates ``image'' to be in disk) to create
space for the image in disk and then stores a {\it reference} to the space in pmem at an
offset of $4$ from the start of the \texttt{addr} variable. The \texttt{setPlace()} function
does a similar thing but uses the pmem allocator (since Listing~\ref{image_disk} annotates
``place'' to be in pmem). The corresponding \texttt{getImage()} and
\texttt{getPlace()} do the appropriate gets.

\subsection{Manual Tagging}
\label{manual_tag}
In manual tagging, the object fields are manually
tagged with the intended storage types.
{\it Multiple tags can be
associated with the same field}.
At run time, the storage
type is decided dynamically based on the available storage space.
For example, for a field with tags of ``@pmem'' and ``@disk'',
after putting the other fields that {\it must} be stored in pmem
(as specified by the user),
if there is enough space in pmem, the field will be stored
in pmem, else it will be stored on disk.
%That is, for fields with multiple possibilities,
%t will be stored in the first location (in the same order as specified in the annotation)
%that can fit the field.
Automatic data promotion/demotion is supported when multiple
tags are defined. For example, the field might initially
be stored in pmem but at a later point of time be shunted
out to disk due to some other field that {\it must} be stored in pmem.
The major disadvantage of this approach is that the user has the non-trivial job of
understanding the usage pattern of the object fields.
\if 0
\begin{lstlisting}[language=nvm, caption=Sample annotation for manual tagging, label=sample]
object person {
            @pmem_ int age;
            /* prefer pmem, else disk */
    @pmem_, @disk_ byte[10000] image; 
            @pmem_ string place;
}
\end{lstlisting}
\fi

\subsection{Profiled Tagging}
\label{profile_tag}
\begin{figure}
    \centering
    \includegraphics[width=\linewidth]{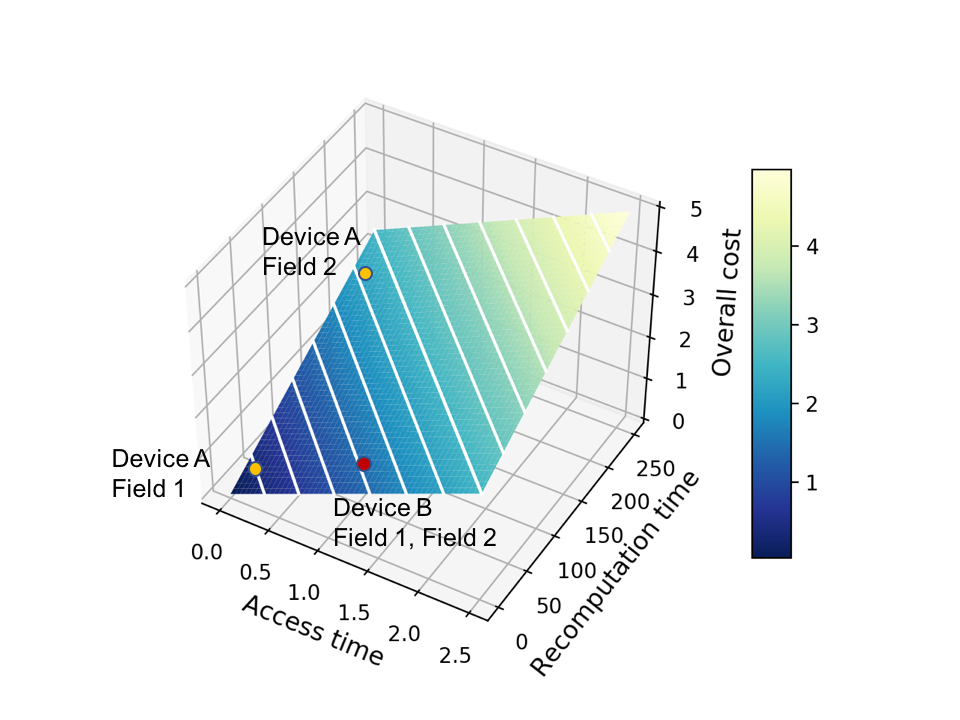}
    \caption{Device selection for 2 different fields in a simulation.}
\label{fig:ilp}
\end{figure}
As opposed to manual tagging, in profiled tagging,
each application has to be profiled {\it a priori}
by running the application on representative
data sets. The profiled data is used to determine the storage type of
the fields. We define the following terms to better formulate the problem:
\begin{itemize}
\item $C$ : Access Time Matrix where $C_{ij}$ corresponds to the access time of field
    $i$ in device $j$. Access time is the total time for field
    access (read/write) on the device. If a device,
    such as SSD/HDD, does not support byte
    addressability, extra serialization/deserialization cost adds to the value.
    {\it Serialization cost} is the cost of serializing a memory
    resident data structure for
    writing to a block device; de-serialization cost is the inverse cost
    of reading a serialized version of a data structure from disk and converting
    it into a memory resident data structure.

\item $F$ : Frequency vector where $F_i$  corresponds to access frequency of a
    field $i$ during a benchmark run.

\item $S$: Storage space vector where $S_j$ corresponds to total storage
    capacity of device $j$ in bytes

\item $R$: Recomputation Time Matrix where $R_{ij}$ corresponds to the time required 
    to recompute the field $i$ (recover the data's {\it pre-failure status}) from device $j$. 
    \if 0
    Recomputation means the extra computation needed to recover the data's 
    {\it pre-failure status} after a program or system failure occurs that causes memory loss.
    \fi
%    Recomputation Time Matrix where $R_{ij}$ corresponds to the time required to
%    recompute the field $i$ from device $j$. If the
%    application had to restart due to failure, non durable fields have to be
%    recomputed again.

\item $P$: Failure probability vector where $P_j$ corresponds to the failure
    probability of device $j$.
    \if 0
    In real world the failure probability can be related with
    various environment factors including network quality, power outage, etc. It
    may be considered same across devices in situations where environment
    caused failures are much more frequent than device caused failures.
    \fi

\item $B$: Field size vector where $B_i$ corresponds to size in bytes required
    to store the field.

\item $X$: Total number of objects to be stored.

\item $a_{ij}$: Binary variable that indicates if field $i$ is stored in device
    $j$.
    \if 0
    This is 1 if field $i$ is stored in device $j$ and 0 otherwise.
    \fi

\end{itemize}

We use an Integer Linear Programming (ILP) formulation to find the
best storage type for each object field as suggested by the collected
profiled data. The optimized object storage cost for all objects,
$\text{\it TotalObjectStorageCost}$, can be defined as the following
minimization problem:
\begin{equation}
\abovedisplayskip 5pt
\belowdisplayskip 4pt
\begin{aligned}
    \text{minimize} \sum_j \sum_i (F_i C_{ij} a_{ij} +  F_i  R_{ij}  P_j a_{ij} ) & & \\
    \text{subject to } (X\sum_{i}B_i a_{ij}) < S_j, \forall j & &
\end{aligned}
\label{eqn:total_cost}
\end{equation}
The first term ($F_i C_{ij} a_{ij}$) is the cost to access field $i$ on device $j$ 
under normal program execution {\it without failure},
i.e., the product of the field's access time and access frequency on the device.
The second term ($F_i  R_{ij}  a_{ij} P_j$) is the cost of accessing field $i$
on device $j$ {\it under failure}, i.e., the product of
the field's access frequency,
the field's recomputation time under failure, and the probability of failure
on the corresponding device.

%The first term ($F_i C_{ij} a_{ij}$) in Equation~\ref{eqn:total_cost} refers to the access time
%of field $i$ when stored on device $j$ while
%the second term ($F_i  R_{ij}  a_{ij} P_j$) refers to the recomputation time of a field $i$ if the field is not
%durable. This is minimized subject to the condition that
%that the total storage capacity of no device is exceeded.

Let us consider a simple scenario where an object containing 2 fields
(Field 1 and Field 2) is being processed in a simulated environment
with 2 different storage devices (Device A and Device B).
An iterative computation task is applied to each field of data,
where number of iterations indicates complexity of computation
(e.g., assume that each field stores a vector of real numbers, at each iteration a matrix
multiplication is applied to the data that transforms it to a new vector and
stores it back to the device). Suppose, Device A is a DRAM and Device B is a PMEM,
a single iteration of computation is applied to Field 1 whereas 10 iterations of 
computation are applied to Field 2.
As shown in Fig.~\ref{fig:ilp}, given empirical values of access time
(e.g., 0.1~us for DRAM~\cite{latency:url}, 1~us for pmem,
i.e., 10~times higher access latency
than DRAM) and a failure rate of 1\%, 
the recomputation time for Device A ($R_{1A}, R_{2A}$) 
is proportional to the complexity of computation on a failure that 
results in memory loss, whereas the recomputation time for 
Device B ($R_{1B}, R_{2B}$) remains constant. In this example, 
the optimal device  choice is PMEM due to its lower recomputation cost on failure.
Since this is based on a simulation, further work is needed to obtain
numbers from more experiments on actual NVDIMM hardware. It is quite possible
that the surface as shown in Fig.~\ref{fig:ilp} might not be as smooth.

Since the profile data for any application is specific to the data sets on which
the profiling was performed, a database of the properties of the data sets and the
corresponding profiled data can be maintained. This would allow {\it estimating}
profiled data for unseen data sets using standard prediction techniques. If the 
new data sets are not very different than the data sets on which the profiling
was run, such prediction could save reduce time spent in profiling.

% Let us consider a simple scenario where an object of 2 fields (i.e., Field 1,
% Field 2) is being processed in a simulated environment with 2 different storage
% devices (i.e. Device A, Device B), as demonstrated in Fig.~\ref{fig:ilp}. Given
% an empirical failure rate $P$ that is indifferent between devices, the $(C_{ij},
% R_{ij}, Cost_{ij})$ falls on a hyperplane, where $C_{ij}$ and $R_{ij}$ are
% empirical values of access costs and recomputation costs of field $i$ on device
% $j$ (available from benchmark tests) and $Cost_{ij}$ is the corresponding overall
% cost. We show qualitatively that the optimal device choice $a_{ij}$ for each
% field can be determined by comparing the $Cost_{ij}$ of each device given that
% the storage constraints are met.

% Suppose that device A is a DRAM and device B is a PMEM, we make a reasonable
% assumption that the access cost $C_{iA} < C_{iB}$ stands for any field $i$ in
% the input data. In one case, field 1 is processed with simple operations
% so that recomputation costs $R_{1A}$, $R_{1B}$ are low on both devices, in this
% case the device A will be preferred due to the lower access cost
% (Fig.~\ref{fig:device_a}). In another case, field 2 is processed with a series
% of complex operations where recomputation costs are high upon failures (i.e.
% $R_{2A} >> R_{2B}$), then the impact of recomputation cost will surpass that of
% access cost, in this case device B will be preferred for field 2 considering the
% overall costs (Fig.~\ref{fig:device_b}).

\subsection{End-to-End Work flow}
We complete the design section with an overview of the end-to-end work flow.
To start off, the traditional class definition is tagged
either manually (Section~\ref{manual_tag}) or
via application profiling (Section~\ref{profile_tag}).
This tagging is used to generate a new class definition
that has APIs (Listing~\ref{generated_apis})
to read and write the fields of the class.
%This is in addition to the
%functions declared in the original class definition,
Finally, the generated class, instead of the original class, is instantiated
in the application code (Listing~\ref{allocator}), and
the generated APIs are used
to read/write the fields of this new, {\it durable} version of the class.
\begin{lstlisting}[language=nvm, caption=Using the generated durable class, label=allocator]
DurablePerson newPerson = new DurablePerson();
/* addr = newPerson.handler();
   addr is internally managed */
newPerson.setAge(10);
newPerson.setImage(byte[1000]{...});
newPerson.setPlace("USA");
newPerson.setName("BOB");
\end{lstlisting}
The proposed tiered storage layout is further extended to {\it durable collections}
such as lists, maps, arrays, and trees.
Durable collections provide durable implementations of the corresponding
Java collections in a transparent way 
and these collections can be used through their
GET/SET/DELETE APIs without knowing their underlying storage layout. 
%The objects are linked via their handler references (object address) internally.

\section{Evaluation}
\label{evaluation}

We considered the following layouts for our evaluation:
\begin{itemize}
\item NO-PMEM refers to a storage layout where all fields needed
for computation are loaded in DRAM.
Current available systems fall in this category.
\item ALL-PMEM refers to a storage layout where all fields lie in PMEM
(Figure~\ref{fig:dram_pmem} and Listing~\ref{image_pmem}).
\item SELECT-PMEM refers to a storage layout where selected object
fields lie in PMEM while others lie in slow storage (disk)
based on the annotations
(Figure~\ref{fig:dram_pmem_disk} and Listing~\ref{image_disk}).
\end{itemize}
In ALL-PMEM and SELECT-PMEM layouts, the Java heap just contains the holders
to the actual storage in the underlying devices.
Actual data gets into heap only if accessed, else data stays in the storage.
This leads to lower heap usage and fewer garbage collection invocations.
%data is loaded into DRAM only when
%accessed, unlike the NO-PMEM case where data is always loaded into DRAM.
Since NVDIMM hardware are not yet available, we had to emulate NVDIMM by carving
out space from DRAM at \texttt{/dev/pmem} and placing an \texttt{ext4} filesystem
on the device as outlined in the Persistent Memory Wiki~\cite{pmem:url}.

%\subsection{$k$-means Clustering}
%\label{kmeans}
%\noindent
%{\bf $k$-means Clustering:}
\subsection{$k$-means Clustering}
\if 0
The experiments are run on Haswell EP
with 75 E5-2699 v3 cores, running at
2.3~GHz, with 246~GB DDR4-2133 RAM and 1.6~TB SSD accessed over a SATA 3.0 (6.0
Gbps) interface. The machine ran CentOS 7 with JDK8 and
Apache Spark 1.5.0\footnote{future work includes porting to newer version of Spark.}
using ParallelOldGC as the garbage collector. The system memory is capped at
128~GB using a dummy ramfile. Spark was running 20 executors with 4 GB JVM heap and
3 cores.
\fi
\if 0
The {\it spark.storage.memoryFraction} parameter specifies the percentage of
Java heap that is used for Spark's memory cache via {\it cache()} call. Since the
StandardKMeans in ML library uses {\it cache()} call to cache partitions, same call is
also added to PersistentKMeans implementation for consistency.
The {\it spark.storage.memoryFraction} is set to 0.6.
%for experiment 1 and 0.2 for
%experiment 2.
\fi
The input dataset is created using a random generator and data is
stored on SSD. The dataset contains 100 million records each with 12 dimensions.
Buffers are cleared after each run. Tiered Storage layout
experiment uses PMemRDD, implemented using a variant of Apache
Mnemonic~\cite{mnemonic:url}, while by default Spark uses regular RDD.
PMemRDD is the durable implementation of Spark's resilient distributed dataset (RDD)
in which generic objects are persisted into durable memory partitions.
\begin{figure}[!tbp]
    \centering
    \includegraphics[width=0.8\linewidth]{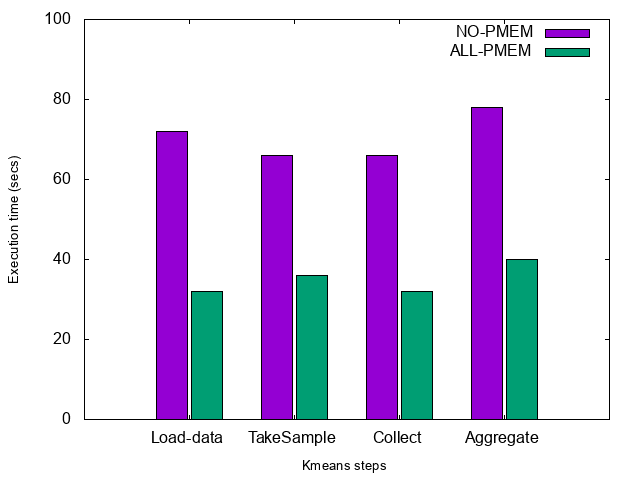}
    \caption{Impact on major steps of $k$-means algorithm. The experiments are run on Haswell EP
with 75 E5-2699 v3 cores, running at
2.3~GHz, with 246~GB DDR4-2133 RAM and 1.6~TB SSD accessed over a SATA 3.0 (6.0
Gbps) interface. The machine ran CentOS 7 with JDK8 and
Apache Spark 1.5.0, running 20 executors with 4 GB JVM heap and
3 cores, and  using ParallelOldGC as the garbage collector.
The system memory was capped at 128~GB using a dummy ramfile.}
    \label{fig:plot_mnemonic_kmeans}
\end{figure}

\begin{table}[t]
\begin{center}
    \resizebox{0.95\columnwidth}{!}{%
\begin{tabular}{||l || c | c | c ||}
    \hline
    GC (Sum of 20 executors) & Default & Tiered Storage & Tiered Storage/Default \\
    \hline\hline
    FullGC (count) & 23 & 12 & 0.522\\
    \hline
    YoungGC (count) & 1310 & 864 & 0.660\\
    \hline
    FullGC pause (sec) & 129.26 & 59.52 & 0.460\\
    \hline
    YoungGC pause (sec) & 225.98 & 158.79 & 0.460\\
    \hline
    Total GC pause (sec) & 355.34 & 218.31 & 0.614\\
    \hline
\end{tabular}
}
\caption{Garbage collection statistics.
Both Full GC and Young GC invocations are reduced substantially.}
\label{tab:spark_gc}
\end{center}
\end{table}

Figure~\ref{fig:plot_mnemonic_kmeans} shows that there is 50\% improvement in
benchmark execution time and in all major steps (X-axis) of the benchmark.
First, there is no extra serialization-deserialization (SerDes) cost
involved when the data is loaded using PMemRDD (ALL-PMEM). 
PMemRDD takes advantage of NVM's byte-addressability feature by 
directly loading objects from pmem each time. 
While in the default case (NO-PMEM) data has to be loaded from input disk
adding SerDes overhead. And second, number of garbage collection invocations 
are 40\% fewer than the default execution, as shown in Table~\ref{tab:spark_gc}.
In ALL-PMEM case, PMemRdd processes non-volatile objects 
directly from pmem without
generating unnecessary temporary objects leading to lower heap pressure
and hence fewer GC triggers, thus improving the overall performance. The impact 
of lower heap pressure is well known and its positive impact 
has been previously reported by Nguyen et al.~\cite{facade:asplos15}, which 
presents compiler optimizations to reduce the number of runtime heap objects.

\subsection{Graph Benchmark}
\label{graph}
For graph benchmarks, we chose Facebook~\cite{snap:facebook}
%and Google+\cite{snap:gplus}
datasets from Stanford Large Network Dataset Collection.
The experiments were run on an Intel NUC, 6th generation i7-6770HQ
processor 2.6~GHZ, Quad Core, 32~GB RAM and 512~GB SSD. The Java heap is set to
4~GB while a separate 5~GB of system memory is reserved as persistent memory.
Both datasets contain a list
of nodes with their friends or circles and their associated features.
%The relations between nodes are indicated by edges.
%Google+ dataset has $>$10 million
%edges while 
The dataset has $>$80,000 {\it edges} (relations between nodes) in the network.
\if 0
One of the most
interesting access pattern on social circle data is a graph search based on node
features or relationships. Some common queries would be ``Return the employer of
all friends who work at $X$ and currently live in city $Y$" or ``Return current
location of all friends who studied at university $X$ and last name is $Y$'.
\fi
Features were selected at random for the experiments. We ran benchmarks on
graph search queries with varying number of search constraints.
For example, a {\it two} field
benchmark in 
Fig.~\ref{fig:plot_mnemonic_fb_load}
translates to ``Return all friends who work at company $X$ and live in city $Y$''.
With SELECT-PMEM, all features used in the search are stored
in pmem while others reside in disk.

\if 0
\begin{figure}
    \minipage{0.23\textwidth}
        \includegraphics[width=\linewidth]{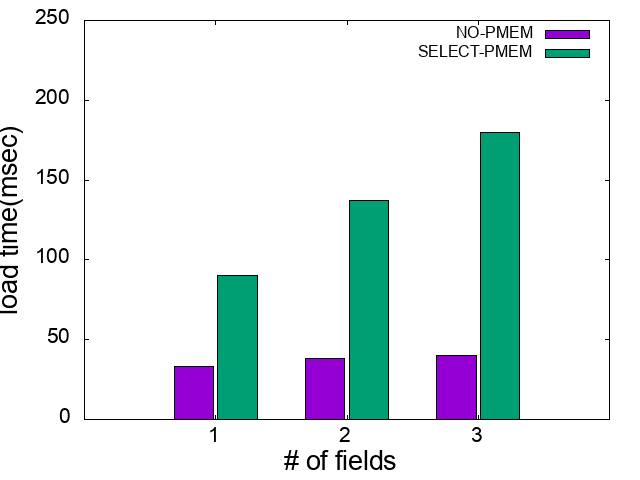}
        \caption{Load time, Graph}
        \label{fig:plot_mnemonic_fb_load}
    \endminipage\hfill
    \minipage{0.23\textwidth}
        \includegraphics[width=\linewidth]{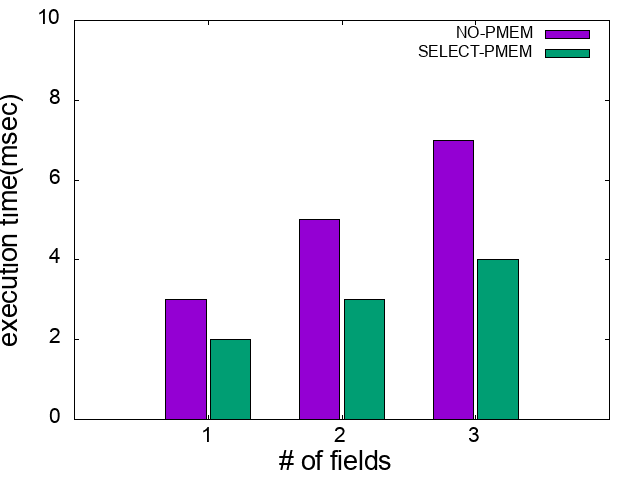}
        \caption{Execution time, Graph}
        \label{fig:plot_mnemonic_fb_exec}
    \endminipage\hfill
\end{figure}
\fi

\begin{figure}
        \includegraphics[width=0.7\linewidth]{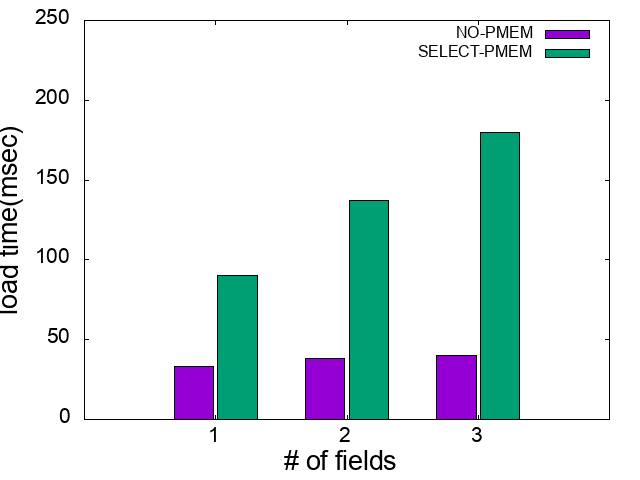}
        \caption{Load time, Graph}
        \label{fig:plot_mnemonic_fb_load}
\end{figure}
\begin{figure}
        \includegraphics[width=0.7\linewidth]{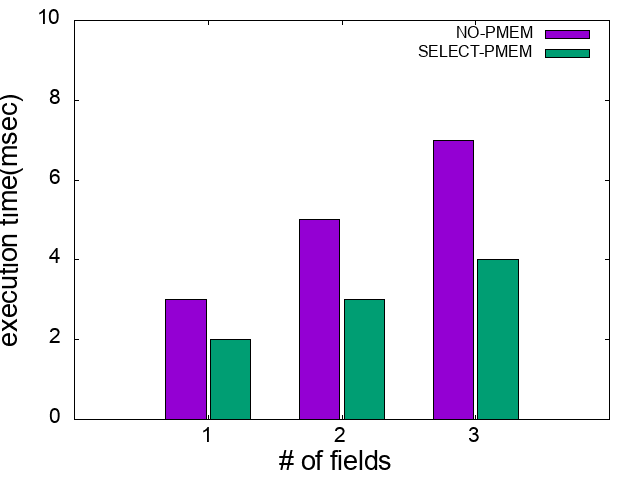}
        \caption{Execution time, Graph}
        \label{fig:plot_mnemonic_fb_exec}
\end{figure}

\if 0
We compare benchmark performance on SELECT-PMEM layout against the default
NO-PMEM layout. In case of SELECT-PMEM layout, only few selected fields of
each node are kept in persistent memory
while other unused fields are kept in disk freeing
persistent memory space.
{\bf XXX which fields are kept in memory and which are not.}
\fi

Loading time in Fig.~\ref{fig:plot_mnemonic_fb_load}
%and Fig.~\ref{fig:plot_mnemonic_gp_load}
refers to the time taken to load data from a
file stored in disk to the target storage (for NO-PMEM, data is loaded into heap
and for SELECT-PMEM, data is loaded into tiered storage).
As expected, the data loading time is higher for
SELECT-PMEM when compared to NO-PMEM. Higher loading cost is due to extra
bookkeeping required to create the tiered storage data layout. However,
{\it Fig.~\ref{fig:plot_mnemonic_fb_exec}
%and Fig.~\ref{fig:plot_mnemonic_gp_exec}
shows that SELECT-PMEM layout shows 30-40\% performance improvement
in execution time over NO-PMEM}.
With increase in search constraints, loading time and execution
time both increase because more data is processed.

\section{Conclusion and Future Work}
\label{future}
%Persistent memory provides byte-addessable access to persistent storage, thus
%enabling programming models that were previously unimaginable.
%This paper presents a novel tiered storage model in which fields of individual data
%structures are stored in different storage layers.
%Our idea is to use profiling to decide the field--to--layer mapping
%in an application specific manner.
%To this end, the
%paper also presents a novel modeling of the problem as an integer linear
%programming problem. Our preliminary results show an execution
%speed up of up to 50\%.
%This is due to a few reasons --- first, it avoids serialization
%and deserialization cost unless absolutely needed and second,
%it reduces the footprint of applications in DRAM.
%This also ensures
%that both main memory and the costlier PMEM are better utilized.
%Future study need to apply this technique to
%additional data intensive systems on machines with real NVDIMM hardware.

This paper presents a tiered storage model in which fields of individual data
structures are stored in different storage layers.
Our idea is to use profiling to decide the field--to--layer mapping
in an application specific manner. The
paper also presents a novel modeling of the problem as an integer linear
programming problem. We show that our tiered model leads to an execution
speed up of up to 50\%.
This is due to a few reasons --- first, it avoids serialization
and deserialization cost unless absolutely needed and second,
it reduces the footprint of applications in DRAM.
This also ensures that both main memory and the costlier PMEM are better utilized.

We are currently implementing tiered storage layout
in Apache Mnemonic~\cite{mnemonic:url} including
durable collections~\cite{mnemonic_collections:url}.
Our immediate goal is to evaluate the proposal using real
NVDIMM hardware. Real hardware is especially important to evaluate
cases where the entire working set cannot fit into pmem and hence data 
has to be moved in and out of pmem.

A possible enhancement is to
augment the traditional distributed storage layer
with a {\it distributed tiered storage layer}
that can be implemented using RDMA
technologies such as InfiniBand~\cite{infiniband}, RDMA over Converged
Ethernet (RoCE)~\cite{roce},
and Internet Wide Area RDMA protocol (iWARP)~\cite{iwarp}.
The distributed implementation inherits all properties of
proposed tiered storage layout and adds scalability
to meet future storage needs.
We believe this paper is a first step towards
several new directions in tiered data layout for data intensive systems. 

\if 0
When datasets are really huge, it has to be
stored across multiple nodes. We are currently exploring different ways to
implement a distributed version of the proposed tiered storage layout spanning
multiple nodes. We already have storage allocators in order to
read and write into distributed systems like Ceph~\cite{ceph:url}.
However, such a layout  does not
leverage memory devices across the network.
One possible enhancement would be to
have a distributed memory layer along with the traditional distributed storage layer.
The distributed memory layer (DRAM and PMEM) can be implemented using RDMA
technologies such as InfiniBand~\cite{infiniband}, RDMA over Converged
Ethernet (RoCE)~\cite{roce},
and Internet Wide Area RDMA protocol (iWARP)~\cite{iwarp}.
All memory to memory communication between systems can be via RDMA
for minimal latency. Such a design avoids expensive network transfers and SerDes
costs involved in data operations; network communication is used only when RDMA
cannot be used. The distributed implementation inherits all properties of
proposed tiered storage layout and adds scalability to meet future storage
needs.

This paper
proposes and demonstrates a novel way of using persistent memory and, in doing
so, contributes to the growing body of work that is on the verge of enabling an
exciting future for software. 
\fi

{\footnotesize \bibliographystyle{acm}
\bibliography{ms}}

%\theendnotes

\end{document}